\documentstyle[twoside,fleqn,epsfig,espcrc2]{article}

\newcommand{\AmS}{{\protect\the\textfont2
  A\kern-.1667em\lower.5ex\hbox{M}\kern-.125emS}}

\title{A New Lattice Action For Studying Topological Charge.}

\author{Pilar Hern\'andez and Raman Sundrum\thanks{Research supported
by NSF-PHYS-92-18167.} 
~~~~~~~~~~~~~~~~~~~~~~~~~~~~~~~~~~~~~~~~~~~~~~~~~~~~~~~~~~~~~~~~~~~~~~~~
~~~~~~~~~~~~~~~~~~~~
Lyman Laboratory of Physics, Harvard University, Cambridge, MA 02138, USA}

\begin{document}

\pagestyle{empty}
         
\begin{abstract}
We review our recent proposal for a new lattice action for non-abelian 
gauge theories which reduces short-range lattice artifacts in the computation
of the topological susceptibility. The standard Wilson action is replaced
by the Wilson action of a gauge covariant interpolation of the original 
fields to a finer lattice. We illustrate the improved behavior of a same-philosophy new lattice action in the $O(3)$ $\sigma$-model in two dimensions. 
\end{abstract}

\maketitle

\section{INTRODUCTION}
     
Field configurations with non-zero topological charge are expected to have a strong influence on the dynamics of asymptotically free theories. The study of these effects however requires non-perturbative techniques and one would
expect that ultimately Monte Carlo methods on the lattice would be best suited to it. The gold-plated observable is the topological susceptibility $\chi_t$, as inspired by the large-$N_c$ analyses. In the continuum it is
given by,
\begin{eqnarray}
\chi_t \equiv \int d^4 x < q(x) q(0) > |_{no\; quarks},
\label{chi}
\end{eqnarray} 
$q(x)$ being the topological charge density,
\begin{eqnarray}
q(x) =  \frac{1}{32 \pi^2} \epsilon_{\mu\nu\rho\sigma} Tr[\;F_{\mu\nu}\;  F_{\rho\sigma}].
\label{contq}
\end{eqnarray}
The topological charge, $Q \equiv \int q(x)$, is an integer if the field 
strength vanishes at infinity or if (euclidean) space-time is compact.   
A continuum analysis shows that the action of any configuration with 
non-zero topological charge must satisfy the following bound, 
\begin{eqnarray}
S \geq \frac{8 \pi^2 |Q|}{g_0^2} 
\label{contbound}
\end{eqnarray}
which is saturated by instantons, $S^{instanton} = 8 \pi^2/g_0^2$. There are several choices for the operator $q(x)$ on the lattice. 
We will deal only with the geometrical definition due to L\"uscher \cite{luscher3}, which gives an integer-valued topological 
charge and does not require renormalization. 

The topological susceptibility in QCD is expected to scale as $(mass)^4$ in the 
continuum limit.
 However, it was found \cite{gksw1} that the Wilson action gives rise to short-range fluctuations with 
non-zero geometrical topological charge and such a small action that they 
overwhelm the contribution of slowly varying fields and can destroy the expected scaling. It is easy to understand how these fluctuations called
dislocations
result
from  the mismatch between the geometrical definition of topological charge
and the Wilson action.
Consider a continuum instanton $A_\mu(y)$ which saturates the 
bound (\ref{contbound}), and discretize it 
on a lattice of spacing $b$,
\begin{eqnarray}
U_\mu(s) \equiv P \exp( i\; \int^{s b+ b \hat{\mu}}_{s b} dy \;A_\mu(y)\;),
\label{dis}
\end{eqnarray}
where $s\; b$ are the sites of the $b$ lattice.
The geometrical definition of topological charge 
assigns a non-zero value 
even to a lattice configuration (\ref{dis}) obtained from very small instantons, of $O(b)$. On the other hand, it is clear that the Wilson action approximates very poorly the continuum 
action for such rough configurations, and 
in fact it turns out to be smaller, strongly violating the bound (\ref{contbound}). 
On the other hand, a  semiclassical continuum analysis of dilute instantons 
indicates that for any theory in which 
the continuum susceptibility is well-defined, there exists an $\alpha < 1$ such that, if the action of topologically non-trivial configurations is always larger 
than $\alpha\; S^{instanton}$, the susceptibility is ultraviolet finite 
\cite{luscher1}. Thus satisfying this minimum bound $S \geq \alpha S^{instaton}$ is a sufficient condition for scaling of the susceptibility. We will show
that our new action satisfies this minimum bound.  

\section{NEW ACTION}

There are two important observations that led us to consider the new action 
proposed in \cite{us1}. The first is that the geometrical 
topological charge assigned to a lattice configuration is just the naive topological 
charge of a {\it continuum} configuration obtained by smoothly interpolating 
the lattice configuration. Then it is clear that if, instead
of using the standard Wilson action of the original lattice configuration, we 
use the continuum action of the interpolated configuration, the continuum bound 
is necessarily satisfied, as first suggested in \cite{gksw}. 
More concretely, in \cite{us2} we described a procedure to obtain a continuum gauge field
$a_{\mu}(y)$ which interpolates any $b$-lattice configuration\footnote{The continuum field when discretized according 
to (\ref{dis}) gives back $U_\mu(s)$ \cite{us2}.}. The interpolation 
is local and gauge covariant, i.e. for a $b$-lattice gauge transformation $\Omega(s)$, 
\begin{eqnarray}
a_\mu[U^\Omega] = a_\mu^\omega[U],
\label{gi}
\end{eqnarray} 
where $\omega$ is a gauge transformation in the continuum.  
A  geometrical
topological charge of the $b$-lattice configuration is defined as the one 
associated to the interpolated field \cite{us2},
\begin{eqnarray}
Q =  \frac{1}{32 \pi^2} \int d^4 y \; Tr[\; \tilde{f}_{\mu\nu}(y) f_{\mu\nu}(y) ]. 
\label{topcharge}
\end{eqnarray}
Now, it is clear that replacing the standard Wilson action by the continuum action of $a_\mu$, i.e.
\begin{eqnarray} 
S^{cont} = \frac{1}{2 g_0^2} \int d^4 y \;Tr[ f_{\mu\nu}(y) f_{\mu\nu}(y)],
\label{contac}
\end{eqnarray}
ensures that the continuum bound is satisfied for the same
reason as it is in the continuum. Notice that 
this is a perfectly gauge invariant action for the lattice field $U_\mu(s)$, by
eq. (\ref{gi}). 
 The only problem with the action (\ref{contac}) is that it is computationally
impractical. 
 
The second and central observation is that using 
(\ref{contac}) is in fact not necessary. An action defined on a finer lattice, with lattice spacing $f$ (we will take $b/f$ to be integer), that approximates
$S^{cont}$ to just enough precision so that the minimum bound is satisfied, will ensure the scaling of the topological susceptibility, as explained above. If the required ratio 
$f/b$ turns out to be 
not too small, the new action will be much easier to compute than (\ref{contac}). 

 More precisely, referring to $x$ as the sites on the $f$-lattice and $s$ as the sites on the 
$b$-lattice ($x_\mu = s_\mu + m_\mu \frac{f}{b}$, $m_\mu = 0,..., \frac{b}{f}-1$), the interpolation procedure 
in \cite{us2} gives a set of link variables $u_\mu[U](x)$, such that
\begin{eqnarray} 
- \frac{i}{f}\; log(u_\mu(x)) = a_\mu(x) + O(f/b^2),
\label{imp}
\end{eqnarray}
 where $a_\mu(x)$ is the 
continuum interpolation discussed above, at point $x$. 
On the $f$ lattice, we can simply choose the standard Wilson action. The partition functional will then have the form, 
\begin{eqnarray}
{\cal Z} =  \int \prod_s {\cal D} U(s) \;e^{- S^f_{wilson}[u[U]]}, 
\label{effact}
\end{eqnarray}
where, $D U$ is the usual Haar measure for non-abelian gauge fields on the $b$-lattice, and the Wilson action in terms of the interpolated link variables $u[U]$ is given by,  
\begin{eqnarray}
S^f_{wilson}[u[U]] = \frac{1}{g_0^2} \sum_x \sum_{\mu\neq\nu} (I - u_{\mu\nu}[x] + h.c. ),
\label{actna}
\end{eqnarray}
with $u_{\mu\nu}[x]$ being the $f$-plaquette variable. Again, this action is gauge invariant, because the functional $u[U]$ is gauge 
covariant \cite{us2} and from eq. (\ref{imp}) it then follows that,
\begin{eqnarray}
S^f_{wilson} = \frac{1}{2 g_0^2} \int d^4 y \;Tr[ f_{\mu\nu}(y) f_{\mu\nu}(y)] + O(f/b).
\nonumber
\end{eqnarray} 
Determining how small the ratio $f/b$ must be to satisfy $S \geq \alpha S^{instanton}$ requires a numerical analysis. In the next section we present a first numerical study of 
this issue in a simplified model in two dimensions. The results encourage
us to believe that $\frac{f}{b}$ need not be very small in order
to recover scaling of $\chi_t$.  

\section{O(3) $\sigma$-MODEL IN 2D}
 
As is well known the O(3) $\sigma$-model in two dimensions shares many similarities with 
Yang-Mills \cite{poly}. 
In the continuum, the model is defined by the action,
\begin{eqnarray}
S= \frac{1}{2 g_0^2} \int d^2 x \sum_\mu \; (\partial_\mu \vec{n} (x))^2,
\label{o3stan}
\end{eqnarray}
where $\vec{n}$ is a 3-component real field satisfying the constraint ${\vec{n}\;} ^2 = 1$. The continuum topological charge in this model 
measures the number of times that space-time wraps around the $\vec{n}$-sphere.
 In a standard lattice treatment the action, $S^b$,  is the naive discretization of (\ref{o3stan}), while
the geometrical topological charge $Q^b$ was first defined on the lattice in \cite{luscher1}. In the continuum, all topologically non-trivial configurations
satisfy $S \geq 4 \pi |Q|/g_0^2$, however, the standard action $S_b$ gives 
rise to dislocations, i.e. configurations with $S^b < 4 \pi |Q^b|/g_0^2$ \cite{luscher1}.
These artifacts disappear when the action is  
improved along the lines of the previous section. The new action, $S^f$, is
defined
on a finer lattice, where the spin variables are obtained from a geodesic
interpolation of the original variables (see \cite{us1} for explicit formulae).
This new action then satisfies     
$S^{f} \geq \frac{4 \pi |Q^b|}{g_0^2} + O(f/b)$ \cite{us1}.

As a first numerical study of the improvement as a function of the ratio $f/b$, we considered the discretization of  continuum instanton configurations with 
unit topological charge \cite{poly} (which saturate the continuum bound). Such configurations are characterized 
by a radius $r$ and a center that we fix at the center of the 
volume to reduce finite volume effects. 
Figure 1 summarizes our results. It represents the action of the discretized instanton configuration as the radius is varied, normalized to the continuum
one-instanton bound. Generically there is always 
a critical, $r_c \sim O(b)$, below which 
$Q^b = 0$ (vertical line). The continuous curve corresponds to 
the standard action, while the dashed ones correspond
to the improved action for different values of the ratio $f/b$. It is
clear that the standard action is problematic, since for $r > r_c$ the action
is considerably smaller than the continuum bound.  For the new 
action however the continuum bound is satisfied already for a ratio of $f/b$ as large as $1/2$! Furthermore, as the instanton is shrunk to sizes of $O(b)$, a small barrier develops, separating the 
$Q^b =0$ and $Q^b = 1$ sectors. This is expected because the  interpolation of a discretized
instanton of size $O(b)$ is generically no longer an instanton (and it is
further from it as the radius of the original instanton decreases). This is 
why the 
action of configurations with $r > r_c$ increases as $r \rightarrow r_c$ near $r_c$. 

Although the results for the $O(3)$ model are very promising, a 
numerical analysis is needed in the Yang-Mills case to determine the required 
ratio $f/b$ there. If it turns out to be moderate, the action (\ref{actna}) should be practical in MC simulations.

\begin{figure}
\begin{center}
\mbox{\epsfig{file=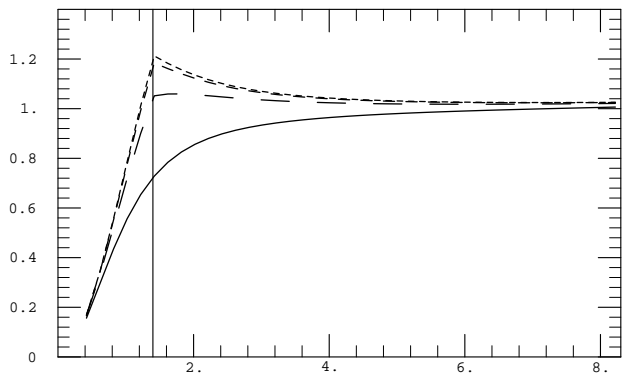,width=7.5cm,bbllx=3.8cm,bblly=9.5cm,bburx=19.0cm,bbury=18.8cm}}    %
\end{center}
 \caption[]{Action of a discretized instanton of the O(3) model (normalized
to $4 \pi/g_0^2$) as
a function of its radius, $r$. The full line is the standard action in a $100 \times 100$ lattice and
the dashed lines correspond respectively to $f/b =1/2, 1/4, 1/6$
(smaller $f/b$, smaller dashing). The vertical line at $r_c \sim 1.4 b$ 
separates the $Q^b = 0, 1$ sectors.}
\label{fig:lattice}
\end{figure}


\begin{thebibliography}{99}

\bibitem{luscher3} M.L\"uscher, Comm.Math.Phys.{\bf 85}(1982)39. 
\bibitem{gksw1}  D.J.R.Pugh, M.Teper, Phys. Lett. {\bf 224B} (1989) 159. M. G\"ockeler et al., Phys. Lett. {\bf 233B} (1989) 192 and references therein.
\bibitem{luscher1} B.Berg, M.L\"uscher, Nucl. Phys. {\bf B190} (1981) 412.
M.L\"uscher, Nucl. Phys. {\bf B200}(1982)61.
\bibitem{us1} P.Hern\'andez, R.Sundrum, hep-lat/9604009.
\bibitem{gksw} M.G\"ockeler et al., Nucl.Phys.{\bf B404}(1993) 839. 
\bibitem{us2} P. Hern\'andez, R. Sundrum, Nucl. Phys. {\bf B472}(1996) 334 and
these proceedings.
\bibitem{poly} E.g. Gauge Fields and Strings, A.M. Polyakov.
\end{thebibliography}
\end{document}